# Arbitrary decomposition of a Mueller matrix

JOSÉ J. GIL[1*], IGNACIO SAN JOSÉ[2]

[1]*Department of Applied Physics, University of Zaragoza, Pedro Cerbuna 12, 50009 Zaragoza, Spain*
[2]*Instituto Aragonés de Estadística, Gobierno de Aragón, Bernardino Ramazzini 5, 50015 Zaragoza, Spain*

*Corresponding author:* ppgil@unizar.es

**Mueller polarimetry involves a variety of instruments and technologies whose importance and scope of applications are rapidly increasing. The exploitation of these powerful resources depends strongly on the mathematical models that underlie the analysis and interpretation of the measured Mueller matrices and, very particularly, on the theorems for their serial and parallel decompositions. In this letter, the most general formulation for the parallel decomposition of a Mueller matrix is presented, which overcomes certain critical limitations of the previous approaches. In addition, the results obtained lead to a generalization of the polarimetric subtraction procedure and allow for a formulation of the arbitrary decomposition that integrates, in a natural way, the passivity criterion.**

Polarimetry constitutes today a very dynamic area in science and engineering that involves powerful measurement techniques widely exploited for the study and analysis of great variety of material samples. Consequently, the mathematical characterization of the polarimetric properties of material media has a capital interest because it provides tools for the analysis and interpretation of experimental measurements in both industrial and scientific environments. The appropriate framework for the mathematical representation of linear polarization interactions is given by the Stokes-Mueller formalism. Mueller matrices are 4×4 real matrices that perform the linear transformation from the Stokes parameters of the incoming state of polarization to the outgoing one. The physical nature of such interactions imposes certain restrictions that are reflected in the fact that the set of Mueller matrices is constituted by a specific subset of real 4×4 matrices.

In analogy to the fact that a Stokes vector can be *pure* (fully polarized) or not, Mueller matrices that preserve the degree of polarization of totally polarized input states (that is, transform any totally polarized Stokes vector into a totally polarized Stokes vector) are called *pure, nondepolarizing*, or *Mueller-Jones* matrices. Thus, Mueller matrices can be pure or not depending on their structural features [1].

The Mueller-Stokes transformations are determined by an ensemble average (a convex sum) of basic pure transformations (*ensemble criterion*) [2,3], each one characterized by a well-defined pure Mueller matrix. This feature leads to the *covariance criterion* that was mathematically formulated by Cloude [4] and, independently, by Arnal [5], through the nonnegativity of the four eigenvalues of the covariance matrix $\mathbf{H}$ associated with a given Mueller matrix $\mathbf{M}$ (thus providing four *covariance inequalities* to be satisfied by $\mathbf{M}$).

A complementary criterion refers to passivity and implies that the action of the medium does not amplify the intensity of the electromagnetic wave interacting with it. More specifically, the assumption of the ensemble criterion entails the necessity that a physically realizable Mueller matrix is susceptible to be expressed as a convex combination of pure and passive Mueller matrices [6].

The main aim of this work is the formulation, in the most general form, of the *arbitrary decomposition* of a Mueller matrix into a convex sum of a minimum number of pure Mueller matrices, in such a manner that the new formulation overcomes the limitation of the previous approaches [7-13] where the Mueller matrices of all parallel components have to be normalized to have equal values for their mean intensity coefficients (defined below), and therefore opens strongly the scope of its applications. This result also provides the way to express the arbitrary decomposition in terms of passive Mueller matrices in accordance with the passivity criterion and, furthermore, leads to a generalization of the polarimetric subtraction procedure [12].

To simplify further expressions, the partitioned block expression of a Mueller matrix [14] wil be used when appropriate

$$\mathbf{M} = m_{00}\,\hat{\mathbf{M}}, \quad \hat{\mathbf{M}} \equiv \begin{pmatrix} 1 & \mathbf{D}^T \\ \mathbf{P} & \mathbf{m} \end{pmatrix},$$

$$\mathbf{m} \equiv \frac{1}{m_{00}}\begin{pmatrix} m_{11} & m_{12} & m_{13} \\ m_{21} & m_{22} & m_{23} \\ m_{31} & m_{32} & m_{33} \end{pmatrix}, \quad (1)$$

$$\mathbf{D} \equiv \frac{(m_{01},m_{02},m_{03})^T}{m_{00}}, \quad \mathbf{P} \equiv \frac{(m_{10},m_{20},m_{30})^T}{m_{00}},$$

where the superscript *T* indicates transpose, $m_{00}$ is the *mean intensity coefficient* (MIC) (i.e. the *transmittance* or *gain* [15-19] of $\mathbf{M}$ for input unpolarized light), while $\mathbf{D}$ and $\mathbf{P}$ are the respective diattenuation and polarizance vectors of $\mathbf{M}$ [20]. The absolute values of these vectors are the diattenuation $D \equiv |\mathbf{D}|$ and the polarizance $P \equiv |\mathbf{P}|$ [20,21]. $\hat{\mathbf{M}}$ denotes the normalized form of $\mathbf{M}$ with MIC $\hat{m}_{00} = 1$. It is also worth to recall that, given the peculiar mathematical structure of a Mueller matrix, its transposed matrix $\mathbf{M}^T$ is also a Mueller matrix [22,23].

Let us consider a light beam with a given state of polarization determined by the corresponding Stokes vector $\mathbf{s}$, whose spot size on a material sample covers *n* areas with different deterministic nondepolarizing polarimetric behavior and that the exiting light pencils are incoherently recombined, so that the state of polarization of the whole outgoing beam is represented by a Stokes vector $\mathbf{s}'$. Thus, the total intensity *I* of the incoming light is shared among *n* portions $I_i$ falling on respective elements "*i*" contained in the illuminated area, which are represented by their respective pure Mueller matrices $\mathbf{M}_{Ji}$ (Fig. 1). The polarimetric transformation of the input Stokes vector **s** into the output **s′** is given by

$$\mathbf{s}' = \sum_{i=1}^{n}\left(\frac{I_i}{I}\mathbf{M}_{Ji}\mathbf{s}\right) = \left(\sum_{i=1}^{n}k_i\mathbf{M}_{Ji}\right)\mathbf{s} = \mathbf{M}\,\mathbf{s},$$

$$\mathbf{M} \equiv \sum_{i=1}^{n}k_i\mathbf{M}_{Ji} = \sum_{i=1}^{n}k_i m_{00i}\hat{\mathbf{M}}_{Ji},\quad k_i \equiv \frac{I_i}{I},\quad \sum_{i=1}^{n}k_i = 1. \quad (2)$$

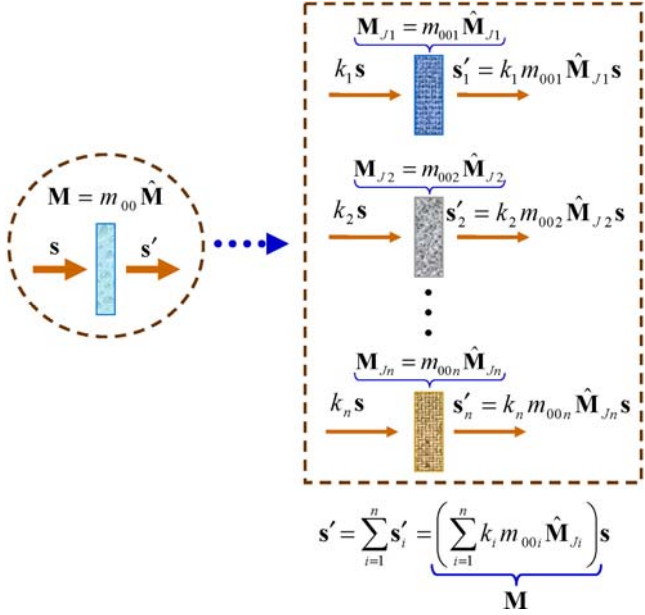

**Fig. 1**. A uniform field with Stokes vector **s** falls on a sample composed of a parallel combination of pure elements characterized by their respective pure Mueller matrices $\mathbf{M}_{Ji} = m_{00i}\hat{\mathbf{M}}_{Ji}$. The portion of intensity falling on each parallel component is given by the respective cross-section $k_i$. The respective outgoing pencils, with Stokes vectors $\mathbf{s}'_i$ are then incoherently recombined into a whole beam with associated Stokes vector $\mathbf{s}'$. The resulting Mueller matrix **M** is given by a convex sum of $\mathbf{M}_{Ji}$ with coefficients $k_i$.

The study of the physically realizable decompositions of **M** into sums of Mueller matrices relies on the statistical nature of depolarizing Mueller matrices [24], which becomes clear when the elements $m_{ij}$ $(i,j=0,1,2,3)$ of **M** are rearranged into the so-called *covariance matrix* **H**, defined as [4,5]

$$\mathbf{H}(\mathbf{M}) = \frac{1}{4}\sum_{i,j=0}^{3}m_{ij}(\boldsymbol{\sigma}_i\otimes\boldsymbol{\sigma}_j), \quad (3)$$

where $\boldsymbol{\sigma}_i$ are the Pauli matrices (arranged in the order commonly used in polarization optics)

$$\boldsymbol{\sigma}_0 = \begin{pmatrix}1&0\\0&1\end{pmatrix},\; \boldsymbol{\sigma}_1 = \begin{pmatrix}1&0\\0&-1\end{pmatrix},\; \boldsymbol{\sigma}_2 = \begin{pmatrix}0&1\\1&0\end{pmatrix},\; \boldsymbol{\sigma}_3 = \begin{pmatrix}0&-i\\i&0\end{pmatrix}. \quad (4)$$

**H** is positive-semidefinite, that is, the four eigenvalues of **H** are nonnegative. Conversely, the elements of **M** can be expressed as follows as functions of **H**

$$m_{ij} = \mathrm{tr}\left[(\boldsymbol{\sigma}_i\otimes\boldsymbol{\sigma}_j)\mathbf{H}\right]. \quad (5)$$

The explicit expressions for $\mathbf{H}(\mathbf{M})$ and $\mathbf{M}(\mathbf{H})$ can be found in [1,25].

Since **H** is a positive semidefinite Hermitian matrix [4], it can be diagonalized as

$$\mathbf{H} = \mathbf{U}\mathrm{diag}(\lambda_1,\lambda_2,\lambda_3,\lambda_4)\mathbf{U}^{\dagger}, \quad (6)$$

where $\lambda_i$ are the four non-negative eigenvalues of **H**, taken in decreasing order $(0\leq\lambda_4\leq\lambda_3\leq\lambda_2\leq\lambda_1)$. The columns $\hat{\mathbf{u}}_i$ $(i=0,1,2,3)$ of the 4×4 unitary matrix **U** are the respective unit, mutually orthogonal, eigenvectors of **H**.

Therefore, **H** can be expressed as the following convex linear combination of four rank-1 covariance matrices that represent respective pure systems

$$\mathbf{H} = \sum_{i=1}^{r}\frac{\lambda_i}{m_{00}}\mathbf{H}_{Ji},\; \mathbf{H}_{Ji} \equiv m_{00}(\hat{\mathbf{u}}_i\otimes\hat{\mathbf{u}}_i^{\dagger}),\; m_{00} = \mathrm{tr}\,\mathbf{H}. \quad (7)$$

where $r\equiv\mathrm{rank}(\mathbf{H})$. Hereafter, when appropriate, pure covariance matrices and pure Mueller matrices will be denoted as $\mathbf{H}_J$ and $\mathbf{M}_J$ respectively.

This (*Cloude decomposition* [4], or *spectral decomposition*) can be written in terms of the corresponding Mueller matrices by means of the following convex sum

$$\mathbf{M} = \sum_{i=1}^{r}\frac{\lambda_i}{m_{00}}\mathbf{M}_{Ji},\; (\mathbf{M}_{Ji})_{00} = m_{00} = \mathrm{tr}\,\mathbf{H}, \quad (8)$$

where all pure Mueller matrices $\mathbf{M}_{Ji}$ have equal MIC, equal to $m_{00}$.

Prior to establish the expressions for the general decomposition of a depolarizing **M** in terms of a minimum number of pure incoherent components of **M**, let us recall that it has been shown that such number is given by $r\equiv\mathrm{rank}(\mathbf{H})$ [8,12].

While the components of the spectral decomposition are defined from the respective eigenvectors $\mathbf{u}_i$ of **H** with nonzero eigenvalue, any Mueller matrix also admit the so-called *arbitrary decomposition* [8,12] (hereafter *homogeneous arbitrary decomposition*)

$$\mathbf{M} = \sum_{i=1}^{r}p_i\mathbf{M}_{Ji} = \sum_{i=1}^{r}p_i m_{00}\hat{\mathbf{M}}_{Ji},$$

$$(\hat{\mathbf{M}}_{Ji})_{ts} = \mathrm{tr}\left[(\boldsymbol{\sigma}_t\otimes\boldsymbol{\sigma}_s)(\hat{\mathbf{w}}_i\otimes\hat{\mathbf{w}}_i^{\dagger})\right],$$

$$p_i = \frac{1}{m_{00}\sum_{j=1}^{r}\frac{1}{\lambda_j}\left|(\mathbf{U}^{\dagger}\hat{\mathbf{w}}_i)_j\right|^2},\quad \sum_{i=1}^{r}p_i = 1, \quad (9)$$

where the subscripts $t,s$ run the elements of $\mathbf{M}_{Ji}$, and $\hat{\mathbf{w}}_i$ $(i=1,...,r)$ is a set of $r$ independent unit vectors belonging to the image subspace of **H** [denoted as $\mathrm{im}(\mathbf{H})$] [12]. Note that when $\hat{\mathbf{w}}_i = \hat{\mathbf{u}}_i$ ($\hat{\mathbf{u}}_i$ being the unit eigenvectors of **H** with nonzero eigenvalue), then the arbitrary decomposition adopts the particular form of the spectral decomposition. The detailed demonstration of the homogeneous arbitrary decomposition can be found in [12]. Note that the denominator of the expression of $p_i$ in (9) can also be expressed as

$$m_{00}\sum_{j=1}^{r}\frac{1}{\lambda_j}\left|(\mathbf{U}^{\dagger}\hat{\mathbf{w}}_i)_j\right|^2 = m_{00}(\hat{\mathbf{w}}_i^{\dagger}\mathbf{H}^{-}\hat{\mathbf{w}}_i), \quad (10)$$

where $\mathbf{H}^{-}$ is the pseudoinverse of **H** defined as $\mathbf{H}^{-} = \mathbf{U}\mathbf{D}^{-}\mathbf{U}^{\dagger}$, $\mathbf{D}^{-}$ being the diagonal matrix whose $r$ first diagonal elements are $1/\lambda_1, 1/\lambda_2,...,1/\lambda_r$ and the last $4-r$ elements are zero.

Decompositions (8) and (9) have been formulated for the case where all pure components have equal MICs, equal to $m_{00}$. This exigency should be avoided because the MICs of the pure components can take specific independent values. In fact, as it will be shown below by means of an example, the homogeneous arbitrary decomposition (9) is not always physically realizable in terms of passive Mueller matrices.

Prior to introduce the generalization of the arbitrary decomposition that contains pure components with arbitrary respective MICs, let us note that by writing a given parallel decomposition (2) in the form (9) and comparing the "*i*" elements appearing in the respective summations in (2) and (9), it follows that

$$k_i\, m_{00i}\, \hat{\mathbf{M}}_{Ji} = p_i\, m_{00}\, \hat{\mathbf{M}}_{Ji} \Rightarrow k_i\, m_{00i} = p_i\, m_{00},$$
$$\left(\sum_{i=1}^{r} k_i = \sum_{i=1}^{r} p_i = 1\right),\tag{11}$$

and therefore the arbitrary decomposition can be expressed in the following generalized form where the components have different MICs denoted as $m_{00i}$ [note that the synthesized expression in (10) is applied]

$$\mathbf{M} = \sum_{i=1}^{r} k_i\, \mathbf{M}_{Ji} = \sum_{i=1}^{r} k_i\, m_{00i}\, \hat{\mathbf{M}}_{Ji},$$
$$k_i = \frac{1}{m_{00i}\left(\hat{\mathbf{w}}_i^{\dagger}\mathbf{H}^{-}\hat{\mathbf{w}}_i\right)},\quad \sum_{i=1}^{r} k_i = 1.\tag{12}$$

Since passivity is a natural feature of experimental samples, the arbitrary decomposition should be performed in terms of passive Mueller matrices. To do so, let us first recall that the necessary and sufficient conditions for a Mueller matrix **M** to be passive are the following [6,26]

$$m_{00}(1+D) \le 1,\quad m_{00}(1+P) \le 1.\tag{13}$$

(Note that, in the case of a pure Mueller matrix, the equality $P = D$ is satisfied [22] and both conditions become a single one). Therefore, the *passive* formulation of the arbitrary decomposition adopts the form

$$\frac{q}{1+X}\hat{\mathbf{M}} = \sum_{i=1}^{r} k_i\, \frac{q_i}{1+X_i}\hat{\mathbf{M}}_{Ji},$$
$$k_i = \frac{1+X_i}{q_i \sum_{j=1}^{r}\frac{1}{\lambda_j}\left|\left(\mathbf{U}^{\dagger}\hat{\mathbf{w}}_i\right)_j\right|^2},\quad \sum_{i=1}^{r} k_i = 1,\tag{14}$$

where

$$X \equiv \max(D,P),\quad X_i \equiv \max(D_i,P_i),$$
$$q \equiv m_{00}(1+X) \le 1,\quad q_i \equiv m_{00}(1+X_i) \le 1.\tag{15}$$

To illustrate the above results, let us consider a parallel composition of two elements, namely, a quarter-wave plate oriented at 0°, with Mueller matrix $\mathbf{M}_R$, and a linear polarizer oriented at 0°, with Mueller matrix $\mathbf{M}_P$ in such a manner that the spot size of the uniform light beam that illuminates this system is shared in such a manner that 1/3 of the intensity falls on the retarder and the remaining 2/3 fall on the polarizer. The composed Mueller matrix **M** is obtained as follows

$$\mathbf{M}_R = \begin{pmatrix} 1 & 0 & 0 & 0 \\ 0 & 1 & 0 & 0 \\ 0 & 0 & 0 & -1 \\ 0 & 0 & 1 & 0 \end{pmatrix},\ \mathbf{M}_P = \frac{1}{2}\begin{pmatrix} 1 & 1 & 0 & 0 \\ 1 & 1 & 0 & 0 \\ 0 & 0 & 0 & 0 \\ 0 & 0 & 0 & 0 \end{pmatrix},$$
$$\mathbf{M} = \frac{1}{3}\mathbf{M}_R + \frac{2}{3}\mathbf{M}_P = \frac{2}{3}\begin{pmatrix} 1 & 1/2 & 0 & 0 \\ 1/2 & 1 & 0 & 0 \\ 0 & 0 & 0 & -1/3 \\ 0 & 0 & 1/3 & 0 \end{pmatrix},\tag{16}$$

so that $m_{001} = 1$, $m_{002} = 1/2$, $m_{00} = 2/3$, $k_1 = 1/3$ and $k_2 = 2/3$. The corresponding homogenous decomposition of **M** takes the form

$$\mathbf{M} = \frac{1}{2}\left(\frac{2}{3}\hat{\mathbf{M}}_R\right) + \frac{1}{2}\left(\frac{2}{3}\hat{\mathbf{M}}_P\right),\tag{17}$$

where the respective coefficients are $p_1 = k_1 m_{001}/m_{00} = 1/2$ and $p_2 = k_2 m_{002}/m_{00} = 1/2$. Note that, since $m_{00} = 2/3 > 1/2$, the polarizer $(2/3)\mathbf{M}_P$ in the homogeneous decomposition does not satisfy the passivity conditions (13) and, therefore, it is not physically realizable. Obviously, given a measured Mueller matrix **M**, the arbitrary decomposition provides infinite specific parallel decompositions of it, and it is the experimentalist, with his experience and knowledge of the problem and its constraints, who can decide which decomposition is more appropriate or plausible for each situation.

As a numerical additional example, let us now consider the application of the arbitrary decomposition (12) to the following experimentally determined Mueller matrix, which corresponds to the reflection (angle of incidence 50⁰) on a steel specimen with surface roughness of 0.256-µm, and with a MgF$_2$ film thickness of 89 nm [27],

$$\hat{\mathbf{M}} = \begin{pmatrix} 1.0000 & 0.1631 & -0.0322 & 0.0802 \\ 0.0083 & 0.4038 & 0.2555 & -0.2158 \\ -0.0026 & 0.4297 & -0.1376 & 0.2016 \\ -0.0116 & 0.0597 & -0.3175 & -0.3690 \end{pmatrix}.\tag{18}$$

The eigenvalues of the covariance matrix **H** associated with $\hat{\mathbf{M}}$ have the nonnegative values

$$\hat{\lambda}_1 = 0.6270,\ \hat{\lambda}_2 = 0.1888,\ \hat{\lambda}_3 = 0.1292,\ \hat{\lambda}_4 = 0.0550,\tag{19}$$

showing that $\hat{\mathbf{M}}$ satisfies the covariance conditions and that $r = 4$ (within the assumed experimental accuracy limits), that is, any arbitrary decomposition should include four parallel components. Other relevant quantities of $\hat{\mathbf{M}}$ that are invariant with respect to *dual-retarder transformations* [28] are $D = 0.1846$, $P = 0.0145$, $P_S = 0.5032$ ($P_S$ being the *degree of spherical purity* [29]), $P_\Delta = 0.5144$ ($P_\Delta$ being the *depolarization index* [21], also called degree of polarimetric purity [8]) and $\det \hat{\mathbf{M}} = 0.1176$. Note that, in this case $D > P$, so that the *passive representative* [6] $\tilde{\mathbf{M}}(\hat{\mathbf{M}})$, (i.e. the Mueller matrix $\tilde{\mathbf{M}}$ proportional to $\hat{\mathbf{M}}$ having the maximal MIC compatible with passivity) is given by

$$\tilde{\mathbf{M}} = \frac{1}{1+D}\hat{\mathbf{M}} = \begin{pmatrix} 0.8442 & 0.1377 & -0.0272 & 0.0677 \\ 0.0070 & 0.3409 & 0.2157 & -0.1822 \\ -0.0022 & 0.3627 & -0.1162 & 0.1702 \\ -0.0098 & 0.0504 & -0.2680 & -0.3115 \end{pmatrix},\tag{20}$$

which admit (among other) an arbitrary decomposition of the form (12), where

$$\mathbf{M}_{J1} = \begin{pmatrix} 1.0000 & 0.0000 & 0.0000 & 0.0000 \\ 0.0000 & 0.3203 & 0.6300 & -0.7074 \\ 0.0000 & 0.8373 & 0.1610 & 0.5225 \\ 0.0000 & 0.4431 & -0.7597 & -0.4760 \end{pmatrix},$$

$$\mathbf{M}_{J2} = \begin{pmatrix} 1.0000 & 0.0000 & 0.0000 & 0.0000 \\ 0.0000 & 0.9744 & -0.1341 & -0.1803 \\ 0.0000 & -0.0825 & -0.9598 & 0.2682 \\ 0.0000 & -0.2091 & -0.2465 & -0.9463 \end{pmatrix},$$

$$\mathbf{M}_{J3} = \begin{pmatrix} 0.6532 & 0.3064 & -0.0605 & 0.1507 \\ 0.2774 & 0.3255 & -0.0377 & 0.5254 \\ 0.1360 & 0.4399 & 0.3158 & -0.1781 \\ -0.1576 & -0.3176 & 0.4571 & 0.1464 \end{pmatrix},$$

$$\mathbf{M}_{J4} = \begin{pmatrix} 0.5527 & 0.3953 & -0.0780 & 0.1944 \\ -0.3577 & -0.4652 & 0.1285 & -0.0195 \\ -0.2096 & -0.2058 & -0.2006 & -0.2583 \\ 0.1679 & 0.0529 & -0.2338 & 0.2758 \end{pmatrix},$$

(21.a)

with

$$k_1 = 0.3713,\ k_2 = 0.2270,\ k_3 = 0.2373,\ k_4 = 0.1644,$$
$$(k_1 + k_2 + k_3 + k_4 = 1).$$

(21.b)

Note that, unlike what would occur with homogeneous arbitrary decompositions, in this particular parallel decomposition the MICs of the components are different, allowing, in addition, that all of them are represented by passive Mueller matrices simultaneously.

Once the arbitrary decomposition has been generalized and even expressed in terms of passive elements, let us revisit the procedure for the polarimetric subtraction and formulate it in the light of this new framework. A given pure component $\mathbf{M}_{J1}$, with associated $\mathbf{H}_{J1} = m_{001}(\mathbf{w}_1^\dagger \otimes \mathbf{w}_1)$, can be considered an arbitrary component of $\mathbf{M}$ if and only if $\operatorname{rank}(\mathbf{H} + \mathbf{H}_{J1}) = \operatorname{rank}\mathbf{H}$ [12]. If this *inclusion* (or *subtractability*) *criterion* is satisfied, then, in accordance with (12), the coefficient $k_1$ corresponding to $\mathbf{H}_{J1}$ in the arbitrary decomposition is given by

$$k_1 = \frac{1}{m_{001}\sum_{j=1}^{r}\frac{1}{\lambda_j}\left|\left(\mathbf{U}^\dagger\hat{\mathbf{w}}_1\right)_j\right|^2} = \frac{1}{m_{001}\left(\hat{\mathbf{w}}_1^\dagger\mathbf{H}^-\hat{\mathbf{w}}_1\right)}. \quad (22)$$

The polarimetric subtraction of $\mathbf{M}_{J1}$ from $\mathbf{M}$ is then performed in the following manner [12]

$$\mathbf{M}_r = \frac{1}{1-k_1}(\mathbf{M} - k_1\mathbf{M}_{J1}). \quad (23)$$

where the rank of the covariance matrix $\mathbf{H}_r$ associated with the resulting matrix $\mathbf{M}_r$ is $r-1$.

If other pure elements are wanted to be consecutively subtracted, the subtraction procedure can be iterated until the difference matrix obtained has rank equal to 1. In each step, the inclusion criterion $\operatorname{rank}(\mathbf{H} + \mathbf{H}_{Ji}) = \operatorname{rank}\mathbf{H}$ should be checked. In addition, as shown in [12], the subtraction of nonpure elements can also be performed and its formulation from the generalized form of the arbitrary decomposition (12) is straightforward.

In summary, unlike the previous approaches, the arbitrary decomposition has been formulated in its most general form, thus allowing one to apply it to any physical or experimental situation and providing the appropriate procedure for the calculation of the coefficients of the parallel components. The new approach has been also expressed in terms of Mueller matrices satisfying the passivity criterion required by natural and man-made samples (except for certain artificial situations [30]). Furthermore, the polarimetric subtraction procedure has been revisited and reformulated in the light of the new generalized arbitrary decomposition framework presented.